\documentstyle[amssymb,prl,aps,epsfig]{revtex}

\begin{document}
\title{Resonant Spin Excitation in an Overdoped High Temperature Superconductor}
\author{H. He$^{1}$, Y. ~Sidis$^{2}$, P.~Bourges$^{2}$, G.D. Gu$^{3}$, A.Ivanov$^{4}$%
, N. Koshizuka$^{5}$, B. Liang$^{1}$,\\
C.T. Lin$^{1}$, L. P. Regnault$^{6}$, E. Schoenherr$^{1}$, and B. Keimer$%
^{1,7}$}
\address{$^{1}$Max-Planck-Institut f\"{u}r Festk\"{o}rperforschung, 70569
Stuttgart,Germany.\\
$^{2}$Laboratoire L\'{e}on Brillouin, CEA-CNRS, CE-Saclay, 91191 Gif sur
Yvette, France.\\
$^{3}$Department of Advanced Electronic Materials, School of Physics,
University of New South Wales, Sydney 2052, Australia.\\
$^{4}$Institut Laue Langevin, 156X, 38042 Grenoble cedex 9, France.\\
$^{5}$SRL/ISTEC, 10-13, Shinonome 1-chome, Koto-ku, Tokyo 135, Japan.\\
$^{6}$CEA Grenoble, D\'{e}partement de Recherche Fondamentale sur la
Mati\`{e}re Conden\'{e}e, 38054 Grenoble cedex 9, France.\\
$^{7}$Department of Physics, Princeton University, Princeton, NJ 08544}
\date{\today}

\twocolumn[\hsize\textwidth\columnwidth\hsize\csname@twocolumnfalse\endcsname

\maketitle

\begin{abstract}
An inelastic neutron scattering study of overdoped Bi$_{2}$Sr$_{2}$CaCu$_{2}$O$_{8+\delta} $ 
(T$_{{\rm c}}=83$ K) has revealed a resonant spin excitation in the
superconducting state. The mode energy is E$_{{\rm res}}=38.0$ meV,
significantly lower than in optimally doped Bi$_{2}$Sr$_{2}$CaCu$_{2}$O$_{8+\delta }$ 
(T$_{{\rm c}}=91$ K, E$_{{\rm res}}=42.4$ meV). This observation, 
which indicates a constant ratio E$_{\rm res}$/$k_B {\rm T}_{{\rm c}}$ $\sim$ 5.4, 
helps resolve a long-standing controversy about the origin of
the resonant spin excitation in high-temperature superconductors.
\end{abstract}

]

A resonant spin excitation \cite
{Rossat91,Mook93,Fong95,Bourges96,revue-cargese,Fong99_1,dai99,Fong99_2}
with wave vector ($\pi ,\pi )$ has recently emerged as a key factor in the
phenomenology of the copper oxide superconductors. In particular, prominent
features in angle-resolved photoemission \cite{Shen97,Norman98,Chubukov99}
and optical conductivity \cite{Munzar99,Carbotte99} spectra have been
attributed to interactions of this bosonic mode with fermionic
quasiparticles. The implications of these observations for the mechanism of
high temperature superconductivity are under intense scrutiny, especially
following suggestions that the spectral weight of the mode (which, at least in
optimally doped YBa$_{2}$Cu$_{3}$O$_{6+x}$, is present only below the
superconducting transition temperature, {\rm T}$_{{\rm c}}$ \cite
{Fong95,Bourges96}) provides a measure of the condensation energy \cite
{Scalapino98,Demler98} or condensate fraction \cite{chakravarty99} of the
superconducting state. Several fundamentally different microscopic
descriptions of the neutron data have been proposed. Some of these \cite
{Fong95,ph} attribute the resonance peak to the threshold of the
particle-hole (ph) spin-flip continuum at $\lesssim 2\Delta _{{\rm SC}}$
where $\Delta _{{\rm SC}}$ is the energy gap in the superconducting state,
others \cite{Chubukov99,coll1,coll2} to a magnon-like collective mode whose
energy is bounded by the gap. Although the starting points of these
calculations are disparate (itinerant band electrons in \cite
{Chubukov99,ph,coll1}, localized electrons in \cite{coll2}), the excitations
corresponding to the neutron peak are described by the same quantum numbers
(spin 1 and charge 0). In a completely different approach \cite{Demler95},
the neutron data are interpreted in terms of a collective mode in the
particle-particle (pp) channel whose quantum numbers are spin 1 and charge
2. The pp continuum that provides the upper bound for the pp resonance in
the latter model is unaffected by superconductivity. Despite the central
significance of this issue, there is still no ``smoking gun'' experiment
selecting the correct theoretical approach.

A careful measurement of the doping dependence of the mode energy, ${\rm E}_{%
{\rm res}}$, can help resolve this issue. In Refs. \cite
{coll1,coll2,Demler95}, the mode is interpreted as a soft mode whose energy
is expected to decrease as a magnetic instability is approached with
decreasing hole content. This is made explicit in an expression derived from
the pp model which predicts that ${\rm E}_{{\rm res}}$ is proportional to
the doping level \cite{Demler95}. The behavior observed in underdoped YBa$%
_{2}$Cu$_{3}$O$_{6+x}$ \cite{revue-cargese,Fong99_1,dai99} is consistent
with that prediction. This alone, however, does not amount to a ``smoking
gun'' because it can also be understood in the framework of the simple ph
pair production model where ${\rm E}_{{\rm res}}\propto {\rm T}_{{\rm c}}$.
In underdoped samples, ${\rm T}_{{\rm c}}$ in turn is monotonically related
to the hole content. Further difficulties derive from ambiguities associated
with the distinction between the normal-state ``pseudo-gap'' in the charge
sector and the true superconducting gap in the underdoped state. These are
mirrored in the spin sector by uncertainties regarding the relationship
between a broad peak observed by neutron scattering in the normal state \cite
{revue-cargese,Fong99_1,dai99} and the sharp resonant peak in the
superconducting state.

The neutron data on underdoped samples therefore do not discriminate clearly
between the very different theories of the resonance peak. Here we report a
neutron scattering study in the overdoped state of Bi$_{2}$Sr$_{2}$CaCu$_{2}$%
O$_{8+\delta }$ where none of these complicating factors are present. In
particular, T$_{{\rm c}}$ is reduced while the hole content keeps
increasing, and the normal-state pseudogap disappears.

To this end, we used an array comprising eight small (individual volumes $%
\sim 0.03$ ${\rm cm}^{3}$), high quality overdoped Bi$_{2}$Sr$_{2}$CaCu$_{2}$%
O$_{8+\delta }$ single crystals grown by the floating-zone method \cite{Gu98}%
. In their as-grown state, the crystals were optimally doped, with T$%
_{c}\sim 91$ K, as was the sample used for our previous neutron study \cite
{Fong99_2}. Using established procedures \cite{Han95}, they were
subsequently annealed at 650$^{{\rm o}}$C under oxygen flow for 200 hours.
(The long annealing time was used as a precaution. Previous studies \cite
{Han95} have shown that 20 hours are sufficient to achieve a homogeneous
oxygen content for identically prepared samples.) Following this, the
individual samples exhibited sharp (width 5-7 K) superconducting transitions
at 83 K, in excellent agreement with prior results \cite{Han95}.
Representative data measured by SQUID magnetometry are shown in the inset to
Fig. \ref{fig4}. The crystals were co-aligned by x-ray Laue diffraction and
mounted in an aluminum holder. The overall mosaicity of the array, $\sim
5^\circ$, was comparable to the angular dependence of the magnetic signal of
the previous study \cite{Fong99_2} and therefore of little consequence for
the signal intensity. However, compared to our previous experiment on a
monolithic, optimally doped single crystal, the imperfect alignment of the
crystal array would introduce additional uncertainties into an absolute
intensity unit calibration which will therefore not be given here. In order
to establish an optimal basis for a comparison of the results on optimally
doped and overdoped samples, the experimental setup precisely duplicated the
one used for the previous study \cite{Fong99_2}. The experiments were
conducted on the triple axis spectrometer IN8 (at the Institut Laue-Langevin
in Grenoble, France) in a focusing configuration with Cu(111) monochromator,
pyrolytic graphite (002) analyzer, and 35 meV fixed final energy. The wave
vector $Q=(H,K,L)$ is given in reciprocal lattice units (r.l.u.), that is,
in units of the reciprocal lattice vectors $a^{\ast }\sim b^{\ast }\sim 1.64$
\AA $^{-1}$ and $c^{\ast }\sim 0.20$ \AA $^{-1}$. In these units, the
in-plane wave vector $(\pi ,\pi )$ is equivalent to ($\frac{h}{2},\frac{k}{2}
$)\ with $h,k$ odd integers. The data were taken with $L$ set close to the
maximum of the intensity modulation due to magnetic coupling of the bilayers
($L=-13.2$ or $L=-14$ for Bi$_{2}$Sr$_{2}$CaCu$_{2}$O$_{8+\delta }$ \cite
{Fong99_2}).

Bi$_{2}$Sr$_{2}$CaCu$_{2}$O$_{8+\delta }$ is a highly complex material with
a multitude of densely spaced phonon branches, not to mention the additional
lattice dynamical complexity due to the incommensurate modulation of the
Bi-O layer. Raw neutron data therefore show a large, featureless background
predominantly due to unresolved single-phonon events. An example is given in
Fig. \ref{fig1}. Building on lessons drawn from work on YBa$_{2}$Cu$_{3}$O$%
_{6+x}$, we have previously established \cite{Fong99_2} how the magnetic
signal can be separated from this background by virtue of its characteristic
energy, momentum, and temperature dependences. Specifically, the magnetic
resonance peak that is the primary focus of the present study gives rise to
a magnetic signal at wave vector $Q=(\pi ,\pi )$ that shows a sharp upturn
below {\rm T}$_{{\rm c}}$ (Refs. \cite
{Fong95,Bourges96,revue-cargese,Fong99_1,Fong99_2}). The first step
therefore involves taking the difference between the measured spectra in the
superconducting and normal states and studying the energy and wave vector
dependence of the enhanced signal. Figs. \ref{fig2} and \ref{fig3} show that
this is confined to a narrow region in energy and wave vector centered at $%
E=38$ meV and $Q=(\pi ,\pi )$, while the background away from this region is
reduced upon cooling. (The temperature dependence of the background becomes
more pronounced at low energies, because the phonon scattering follows the
Bose population factor $(1-\exp (-E/k_{B}T))^{-1}$.) This is precisely the
signature of the magnetic resonance peak observed in YBa$_{2}$Cu$_{3}$O$%
_{6+x}$ and optimally doped Bi$_{2}$Sr$_{2}$CaCu$_{2}$O$_{8+\delta }$.

The data on overdoped and optimally doped Bi$_{2}$Sr$_{2}$CaCu$_{2}$O$%
_{8+\delta }$ (also shown for comparison in Fig. \ref{fig2}) were fitted to
a Gaussian magnetic resonant mode on top of a phonon background whose energy
dependence is determined independently from scans at high temperatures and
from constant-q scans away from $(\pi,\pi)$. The phonon background is
multiplied by the Bose population factor, and the difference between the
Bose factors at low and high temperatures gives rise to the negative signal
at low energies in the difference plots. (The fact that the high temperature
scan was taken at 100 K for the material with ${\rm T_c =91}$ K, and at 90 K
for the one with ${\rm T_c =83}$ K, was taken into account in this analysis
and did not influence the result.) Apart from an overall scale factor, there
are two free parameters in the fit: The intensity of the resonant mode with
respect to the phonon background, and its position. The results of these
fits are shown in Fig. \ref{fig2}. The resonance energies extracted in this
manner for optimally doped and overdoped Bi$_{2}$Sr$_{2}$CaCu$_{2}$O$%
_{8+\delta }$ are $42.4 \pm 0.8$ meV and $38.0 \pm 0.6$ meV, respectively
(95\% confidence limits). If the assumption of equal widths of the resonant
mode in both samples is relaxed, the relative position extracted from the
fits is hardly affected. Likewise, a Lorentzian profile for the resonant
mode also gave the same result within the error. The null hypothesis (no
energy shift) can therefore be ruled out with a statistical confidence well
exceeding 95\%.

Before discussing the implications of these data, we proceed to the second
step in the identification of the resonance peak, namely, the determination
of the onset temperature of the magnetic signal. The temperature dependence
of the peak magnetic intensity, shown in Fig. \ref{fig4}, indeed exhibits
the strong upturn around {\rm T}$_{{\rm c}}=83$ K that characterizes the
resonance peak. Interestingly, this upturn is even sharper here than in the
optimally doped sample. As in optimally doped YBa$_{2}$Cu$_{3}$O$_{7}$
(Refs. \cite{Fong95,Bourges96}) and Bi$_{2}$Sr$_{2}$CaCu$_{2}$O$_{8+\delta }$
(Ref. \cite{Fong99_2}), there is no evidence of magnetic scattering above 
{\rm T}$_{{\rm c}}$ although this determination is limited by the high
nuclear background.

The data shown in Figs. \ref{fig2}-\ref{fig4} are an essential complement to
an extensive data set on the resonance peak in underdoped YBa$_{2}$Cu$_{3}$O$%
_{6+x}$. (Note that some \cite{revue-cargese}, but not all \cite{Fong95},
data on the resonant mode in slightly overdoped YBa$_{2}$Cu$_{3}$O$_{7}$
also exhibit a subtle trend towards lower energies. However, the data of
Ref. \cite{revue-cargese} were taken under conditions different from those
on underdoped and optimally doped YBa$_{2}$Cu$_{3}$O$_{6+x}$, which prevents
an accurate comparison of the resonance energies.) A representative subset 
\cite{revue-cargese,Fong99_1,dai99} is shown in Fig. \ref{fig5} along with
the presently available data on Bi$_{2}$Sr$_{2}$CaCu$_{2}$O$_{8+\delta }$.
While we did not confirm the linear relationship between ${\rm E}_{{\rm res}}
$ and the doping level predicted by the pp model of the resonance peak \cite
{Demler95}, Fig. \ref{fig5} suggests that the parameter controlling ${\rm E}%
_{{\rm res}}$ is actually the transition temperature {\rm T}$_{{\rm c}}$,
with E$_{{\rm res}}$/$k_{B}{\rm T}_{{\rm c}}$ $\sim $ 5.4. Since at least 
in the underdoped regime the superconducting gap does not
scale with ${\rm T}_{{\rm c}}$, this observation is not naturally understood
within the ph scenario either. Along with some aspects of the sharp 
``quasiparticle peak'' observed in photoemission data in the 
superconducting state \cite{Feng00}, the neutron resonance thus appears
to be one of very few spectral features of the superconducting cuprates 
that scale with ${\rm T}_{{\rm c}}$. While this may indicate a smooth 
crossover between a magnon-like collective mode below the ph continuum in 
the underdoped regime and a simple ph pair production scenario in the 
overdoped regime, a quantitative theory of such a crossover has thus far 
not been reported.   

Our result supports the conclusion of a neutron scattering study \cite
{ybco-ni} of 3\%-Ni substituted YBa$_2$Cu$_3$O$_7$ (T$_{{\rm c}}$=80 K). Ni
substitution is known to reduce T$_{{\rm c}}$ but does not affect the hole
content. In YBa$_2$(Cu$_{0.97}$Ni$_{0.03}$)$_3$O$_7$, E$_{{\rm res}}$ is
shifted from 40 meV to $\sim$ 35 meV so that the ratio E$_{{\rm res}}$/T$_{%
{\rm c}}$ is preserved, as observed here for overdoped Bi$_{2}$Sr$_{2}$CaCu$%
_{2}$O$_{8+\delta }$.

In conclusion, we have shown that the energy of the magnetic resonant mode
scales with the superconducting transition temperature in both the
underdoped and the overdoped regimes. This result is important for
current theoretical efforts \cite{Shen97,Norman98,Chubukov99,Munzar99,Carbotte99} 
to develop a unified phenomenology of magnetic and charge spectroscopies of the cuprates.

We gratefully acknowledge discussions with P.W. Anderson, S. Chakravarty, A.
Chubukov, E. Demler, W. Hanke, D. Morr, F. Onufrieva, P. Pfeuty, D. Pines,
S. Sachdev, and S.C. Zhang. The work at Princeton University was supported by the
National Science Foundation under DMR-9809483.

\begin{figure}[tbp]
\caption{ Raw (unsubtracted) spectra at wave vector $Q=(0.5,0.5,-13.2)$ and
temperatures 10K and 100K. The large background is mostly due to a multitude
of unresolved phonons (see text). The error bars are smaller than the symbol
size.}
\label{fig1}
\end{figure}

\begin{figure}[tbp]
\caption{Closed symbols: Difference spectrum of the neutron intensities at T
= 5 K (${\rm < T_c}$) and T = 90 K (${\rm > T_c}$), at wave vector ${\bf Q}
= (0.5,0.5,-13.2)$ for overdoped Bi$_{2}$Sr$_{2}$CaCu$_{2}$O$_{8+\protect%
\delta}$. Open symbols: Data taken under identical conditions in optimally
doped Bi$_{2}$Sr$_{2}$CaCu$_{2}$O$_{8+\protect\delta}$ (Ref. \protect\cite
{Fong99_2}). The bar respresents the instrumental energy resolution, The
solid lines are the results of fits as described in the text. The dashed
line is the resonance energy extracted from fitting the data on optimally
doped Bi$_{2}$Sr$_{2}$CaCu$_{2}$O$_{8+\protect\delta}$.}
\label{fig2}
\end{figure}

\begin{figure}[tbp]
\caption{Difference spectrum of the neutron intensities at T = 5 K (${\rm <
T_c}$) and T = 90 K (${\rm > T_c}$), at energy 38 meV. The bar respresents
the instrumental momentum resolution, the line is a guide-to-the-eye.}
\label{fig3}
\end{figure}

\begin{figure}[tbp]
\caption{ Temperature dependence of the neutron intensity at energy 38 meV
and wave vector ${\bf Q} = (0.5,0.5,-13.2)$. The intensity falls to
background level above ${\rm T_c = 83}$ K. The line is a guide-to-the-eye.
The insert shows a measurement of the uniform magnetization of a sample from
the same batch in an applied field of 10G.}
\label{fig4}
\end{figure}

\begin{figure}[tbp]
\caption{ A synopsis of the resonance peak energy ${\rm E_{res}}$ in
underdoped and optimally doped YBa$_{2}$Cu$_{3}$O$_{6+x}$ (open symbols,
with squares from Ref. \protect\cite{revue-cargese}, circles from Ref. 
\protect\cite{Fong99_1}, and diamonds from Ref. \protect\cite{dai99}) and
optimally doped and overdoped Bi$_{2}$Sr$_{2}$CaCu$_{2}$O$_{8+\protect\delta
}$ (closed symbols, from Ref. \protect\cite{Fong99_2} and present work),
plotted as a function of the superconducting transition temperature ${\rm %
T_{c}}$. The shaded areas indicate measures of (or upper bounds on)
the intrinsic width of the peak. The error bar on the peak
position is of the order of the symbol size. }
\label{fig5}
\end{figure}


\begin{references}
\bibitem{Rossat91}  J. Rossat-Mignod {\it et al.}, Physica C {\bf 185-189},
86 (1991).

\bibitem{Mook93}  H.A. Mook {\it et al.}, Phys. Rev. Lett. {\bf 70}, 3490
(1993).

\bibitem{Fong95}  H.F. Fong {\it et al.}, Phys. Rev. Lett. {\bf 75}, 316
(1995); Phys. Rev. B {\bf 54}, 6708 (1996).

\bibitem{Bourges96}  P. Bourges {\it et al.}, Phys. Rev. B {\bf 53}, 876
(1996).

\bibitem{revue-cargese}  P. Bourges, in {\it The Gap Symmetry and
Fluctuations in High Temperature Superconductors}, edited by J. Bok, G,
Deutscher, D. Pavuna and S.A. Wolf. (Plenum Press, 1998) 349.

\bibitem{Fong99_1}  H.F. Fong {\it et al}., Phys. Rev. B {\bf 61}, 14773
(2000), and references therein.

\bibitem{dai99}  P. Dai {\it et al.}, Science {\bf 284}, 1344 (1999).

\bibitem{Fong99_2}  H.F. Fong {\it et al.}, Nature {\bf 398}, 588 (1999).

\bibitem{Shen97}  Z.X. Shen and J.R. Schrieffer, Phys. Rev. Lett. {\bf 78},
1771 (1997).

\bibitem{Norman98}  J.C. Campuzano {\it et al}., Phys. Rev. Lett. {\bf 83},
3709 (1999); M.R. Norman and H. Ding, Phys. Rev. B {\bf 57}, R11089 (1998).

\bibitem{Chubukov99}  A. Abanov and A.V. Chubukov, Phys. Rev. Lett. {\bf 83}%
, 1652 (1999).

\bibitem{Munzar99}  D. Munzar, C. Bernhard, and M. Cardona, Physica C {\bf %
318}, 547 (1999).

\bibitem{Carbotte99}  J.P. Carbotte, E. Schachinger, and D.N. Basov, Nature 
{\bf 401}, 354 (1999).

\bibitem{Scalapino98}  D.J. Scalapino and S.R. White, Phys. Rev. B {\bf 58},
8222 (1998).

\bibitem{Demler98}  E. Demler and S.C. Zhang, Nature {\bf 396}, 733 (1998).

\bibitem{chakravarty99}  S. Chakravarty and H.K. Kee, Phys. Rev. B {\bf 61},
14821 (2000).

\bibitem{ph}  See, {\it e.g.}, L. Yin, S. Chakravarty and P.W. Anderson,
Phys. Rev. Lett. {\bf 78}, 3559 (1997); A.A. Abrikosov, Phys. Rev. B {\bf 57}%
, 8656 (1998).

\bibitem{coll1}  See, {\it e.g.}, I.I. Mazin and V.M. Yakovenko, Phys. Rev.
Lett. {\bf 75}, 4134 (1995); F. Onufrieva, Physica C {\bf 251}, 348 (1995);
D.L. Liu, Y. Zha and K. Levin, Phys. Rev. Lett. {\bf 75}, 4130 (1995); N.
Bulut and D.J. Scalapino, Phys. Rev. B {\bf 53}, 5149 (1996); A.J. Millis
and H. Monien, Phys. Rev. B {\bf 54}, 16172 (1996); J. Brinckmann and P.A.
Lee, Phys. Rev. Lett. {\bf 82}, 2915 (1999); F. Onufrieva and P. Pfeuty,
cond-mat/9903097.

\bibitem{coll2}  A.V. Chubukov, S. Sachdev, and J. Ye, Phys. Rev. B {\bf 49}%
, 11919 (1994); D.K. Morr and D. Pines, Phys. Rev. Lett. {\bf 81}, 1086
(1998); S. Sachdev, C. Buragohain, and M. Vojta, Science {\bf 286}, 2479
(1999).

\bibitem{Demler95}  E. Demler and S.C. Zhang, Phys. Rev. Lett. {\bf 75},
4126 (1995); E. Demler, H. Kohno, and S.C. Zhang, Phys. Rev. B {\bf 58},
5719 (1998).

\bibitem{Gu98}  G.D. Gu, K. Takamuku, N. Koshizuka, and S. Tanaka, J.
Crystal Growth {\bf 130}, 325 (1998).

\bibitem{Han95}  S.H. Han {\it et al}., Physica C {\bf 246}, 22 (1995).

\bibitem{ybco-ni}  Y. Sidis {\it et al}., Phys. Rev. Lett. {\bf 84}, 5900
(2000).

\bibitem{Feng00} D.L. Feng {\it et al.}, Science {\bf 289}, 277 (2000).
\end{references}
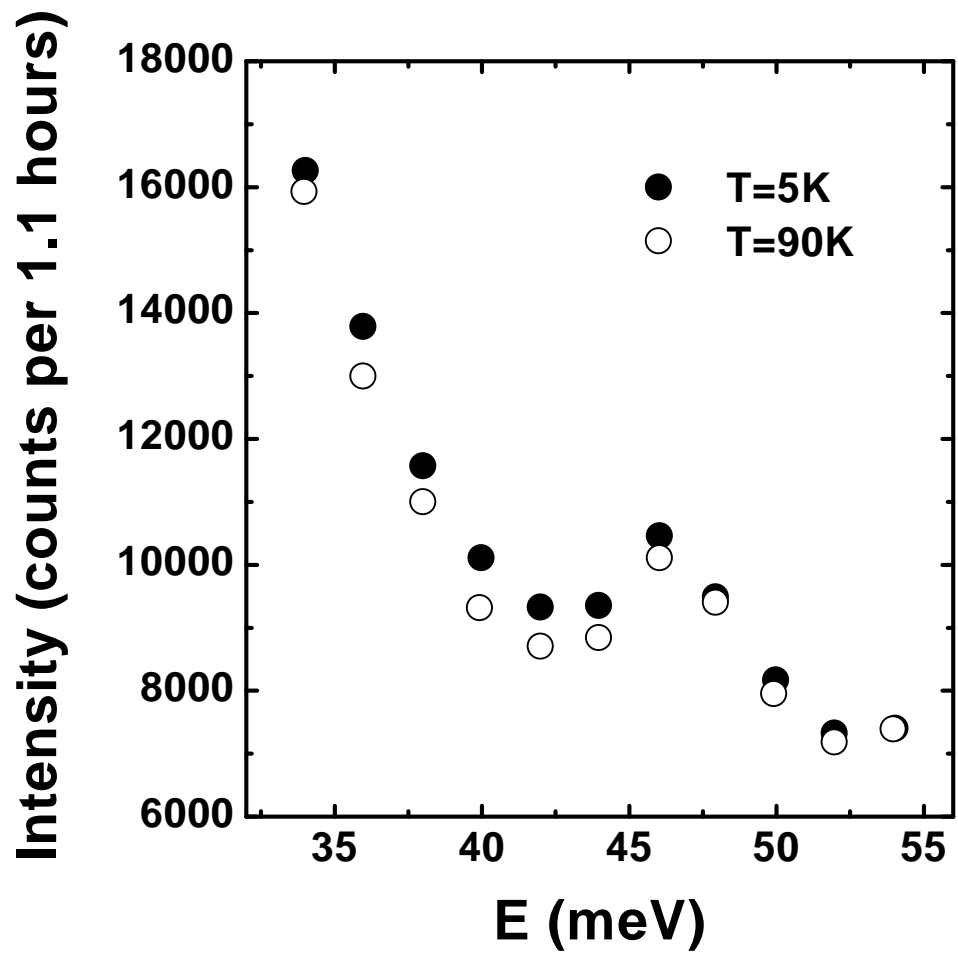
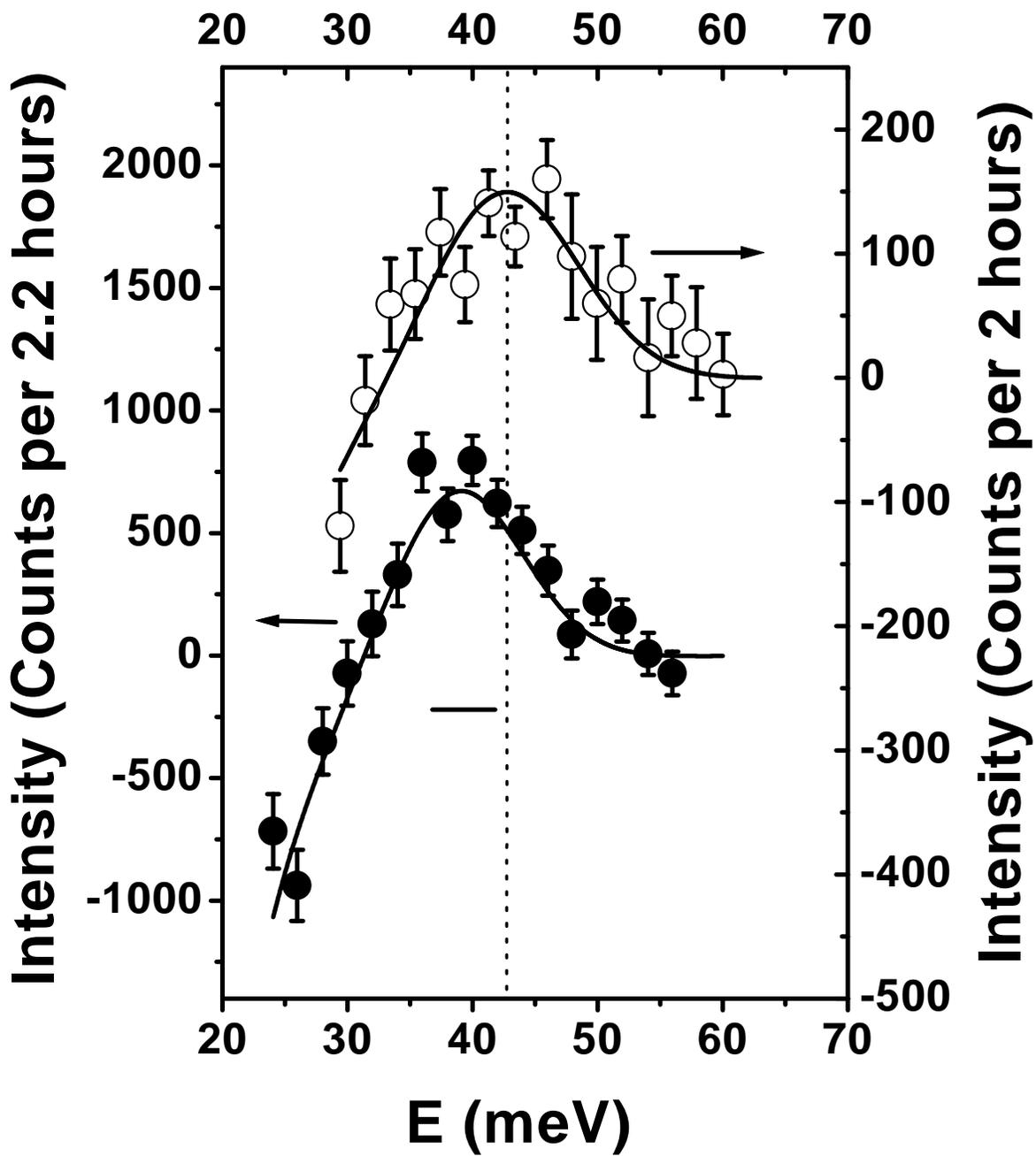
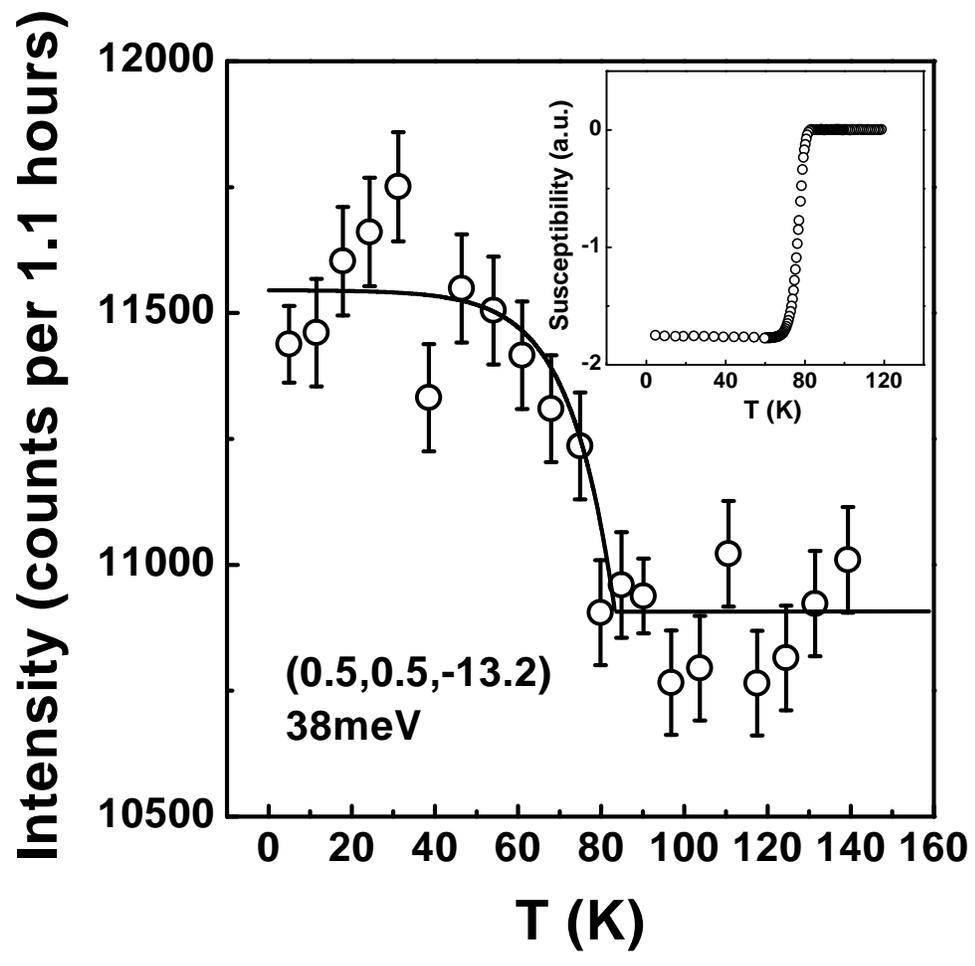
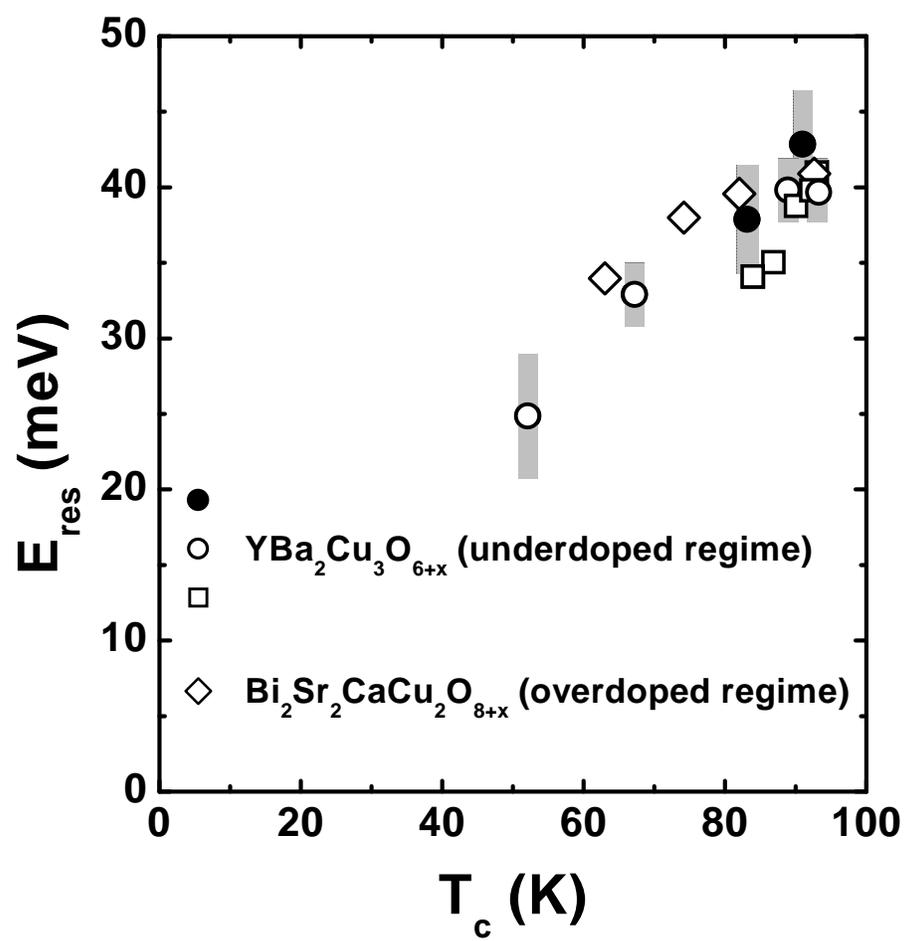
\end{document}